\documentclass[10pt]{article}
\usepackage{titlesec}
\usepackage{amsmath,esdiff}
\usepackage{amssymb}
\usepackage{authblk}

\usepackage[normalem]{ulem}
\usepackage{bbold}
\usepackage{BOONDOX-cal}
\usepackage[colorlinks=true,linkcolor=blue,citecolor=teal,urlcolor=teal]{hyperref}%
\usepackage{geometry}
\usepackage{setspace}
\usepackage{cite}
\usepackage{flushend}
\usepackage{mathrsfs}
\usepackage{hyperref}
\usepackage{graphicx}
\usepackage{cancel}
\usepackage{array,esint}
\usepackage[most]{tcolorbox}
\graphicspath{}
\titleformat{\section}{\fontsize{12}{12}\bfseries}{\thesection}{1em}{}
\twocolumn
\begin{document}
\twocolumn[\begin{@twocolumnfalse}
\title{\textbf{Breaking of universal nature of central charge criticality in \textit{AdS} black holes in Gauss-Bonnet Gravity}}
\author{\textbf{Neeraj Kumar${}^{a*}$, Soham Sen${}^{a\dagger}$ and Sunandan Gangopadhyay${}^{a\ddagger}$}}
\affil{{${}^a$ Department of Astrophysics and High Energy Physics}\\
{S.N. Bose National Centre for Basic Sciences}\\
{JD Block, Sector III, Salt Lake, Kolkata 700 106, India}}
\date{}
\maketitle
\begin{abstract}
\noindent In this paper, we have studied the thermodynamics of Gauss-Bonnet black holes in D-dimensional \textit{AdS} spacetime. Here, the cosmological constant ($\Lambda$), Newton's gravitational constant ($G$) and the Gauss-Bonnet parameter ($\alpha$) are varied in the bulk, and a mixed first law is rewritten considering central charge ($C$) (of dual boundary conformal theory) and its conjugate variable utilising the gauge-gravity duality. A novel universal nature of central charge near the critical point of black hole phase transition in Einstein's gravity has been observed in \cite{mann1}. We observe that this universal nature breaks when such phase transition is considered for black holes in the Gauss-Bonnet gravity. Apart from this, treating the Gauss-Bonnet parameter as a thermodynamic variable as suggested in \cite{kastori} in light of the consistency between first law and the Smarr relation leads to modified thermodynamic volume (conjugate to variable cosmological constant), adding to a new understanding of the Van der Waals gas like behaviour of the black holes in higher dimensional and higher curvature gravity theories. Our analysis considers a general $D$ dimensional background. We have then imposed a greater focus in the analysis of the phase structure of the five dimensional Gauss-Bonnet spacetime. Our analysis also shows that the general universal nature of the critical value of the central charge (which was present in four dimensional \textit{AdS} spacetime), breaks down in case of five dimensional \textit{AdS} spacetime even in the absence of Gauss-Bonnet gravity. This finding indicates the universal nature of the central charge may be a special feature of the four dimensional \textit{AdS} spacetime only.
\end{abstract}
\end{@twocolumnfalse}]
\section{Introduction}
\noindent\footnote{{}\\
{$*$neerajkumar@bose.res.in}\\
{$\dagger$sensohomhary@gmail.com, soham.sen@bose.res.in}\\
{$\ddagger$sunandan.gangopadhyay@gmail.com}}
\noindent Black holes are the most fascinating solutions of the Einstein's field equations in general relativity which have been a part of active area of research since their discovery. Macroscopically, these are very simple objects and need only a few parameters to define completely. Most general black hole is defined by a few parameters, mass (M), charge (Q) and angular momentum (J) \cite{uni1, uni11, uni2, uni3, uni4,  uni5} together with entropy \cite{Bekenstein, Bekenstein2} and Hawking temperature \cite{Hawking, Hawking2, Hawking3}. These variables complete the thermodynamic description of a black hole spacetime. However, very recently, black hole thermodynamics has incorporated cosmological constant ($\Lambda$) as another thermodynamic variable which has given the pressure-volume term in the first law of black hole thermodynamics\cite{kastor,MM, Dolan, Dolan2, Dolan3, Cvetic2, LuPang}. This new paradigm has been named black hole chemistry \cite{revmann1, bhc}. Here, negative cosmological constant has been realized as positive thermodynamic pressure. Hence, black holes in \textit{AdS} (anti-de sitter) spacetime have gained more attention. Mass of the black hole is now interpreted as enthalpy rather than conventional internal energy in this new setting. Also, the new framework has resulted into new interpretations of phase structure of black holes \cite{ext1,ext2,rongPV, gunasekaran, ext3,ext4,ext5}. Most importantly, a correspondence between a black hole spacetime and Van der Waals fluid has been established where black holes are shown to undergo first order small to large black hole phase transition and show similar critical behaviour \cite{ext2, gunasekaran}. Hence, under the discipline of black hole chemistry, black holes have more resemblence to standard thermodynamic systems and provide new explanations to otherwise obscure thermodynamic properties of a black hole.

\noindent Thermodynamic understanding of black holes is also important from the AdS/CFT point of view \cite{mald, gub, revads}. According to this duality correspondence, a black hole in AdS spacetime corresponds to a system described by conformal field theory at finite temperature. The conjecture so far has been exploited to understand many field theory problems \cite{wit, horo}. Varying cosmological constant in AdS spacetime changes the interpretations of boundary theory.  It has been shown that the variation in cosmological constant correspond to change in boundary volume \cite{cv} as well as variation in the central charge of the underlying CFT\cite{karch}. Hence, pressure-volume term in first law in the bulk corresponds to two terms in the holographic first law which is problematic. This issue is resolved in \cite{visser} by making Newton's constant ($G$) a thermodynamic variable in the bulk which allows one to keep the central charge fixed. This formalism results in modification of the first law in bulk which has a variable Newton's constant and its conjugate variable term. Exploiting the \textit{AdS}/CFT dictionary one replaces the variable Newton's constant ($\delta G$) term by variable central charge and its conjugate and the new form of first law is termed as `mixed first law' \cite{mann1}. Most important results of this formalism are the new definition of the volume (conjugate variable of pressure) and phase structure of the black hole. Free energy analysis reveals that at critical point, the central charge is universal and does not depend on $G$ (hence independent of $P$ as well). Very recently, it has been shown in our recent work\cite{nss} that the universality of critical central charge breaks when Born-Infeld nonlinearity \cite{Born} is introduced in black hole solution \cite{nss}. All the results discussed so far are in Einstein's gravity. Hence, it is natural to ask about the validity of these in modified theories of gravity. This paper discusses the simplest extension in the form of Gauss-Bonnet gravity as it also admits black hole solution in AdS spacetime \cite{DL}. Gauss-Bonnet gravity is the simplest extension to Einstein's gravity in the sense that no higher order derivatives than second order appear in the equation of motion. The solutions are non-trivial for spacetime dimensions five and more.
Gauss-Bonnet parameter being dimensionful enters in the first law and Smarr relation \cite{Smarr} for the black hole \cite{kastori}. Thermodynamics of Gauss-Bonnet black holes have been studied extensively in \cite{Gauss1} - \cite{Gauss5} (see also references there in). Our aim in this paper is to understand how the Gauss-Bonnet parameter changes the definition of the volume along with the effects on the phase structure of the black hole especially near critical point when mixed first law is considered.

\noindent The paper is organised as follows. Section 2 discusses thermodynamic properties of the black hole in general $D$ dimensions in Gauss-Bonnet gravity. Section 3 is devoted to the calculations of mixed first law where the new expression for volume of the black hole is derived. Free energy analysis for dimensions five is done in section 4 and the phase transition properties are studied. General dimension dependence of the phase transition is also shown. Section 5 discusses the nature of the critical central charge and its dependence on the Gauss-Bonnet parameter. We summarise our results in section 6.   
\section{Thermodynamics of black holes in Gauss-Bonnet \textit{AdS} spacetime}
\noindent In this section, we review charged black holes in Gauss-Bonnet gravity and its thermodynamic properties in $AdS$ background. The action in general $D$-dimensional \textit{AdS} spacetime reads \cite{DL} 
\begin{equation}\label{1.1}
S=\dfrac{1}{16\pi G}\int d^{D}x\sqrt{-g}\left[R-2\Lambda+\alpha L_{GB}+L(F)\right]~
\end{equation}
where the Gauss-Bonnet Lagrangian density ($L_{GB}$) is of the form
\begin{equation}\label{1.2}
L_{GB}=R^2-4R_{\gamma\delta}R^{\gamma\delta}+R_{\gamma\delta\lambda\sigma}R^{\gamma\delta\lambda\sigma}.
\end{equation}
 In eq.(\ref{1.1}), $L(F)=-F^{\mu\nu}F_{\mu\nu}$ is the Maxwell field Lagrangian with $F^{\mu\nu}=\partial_{\mu} A_{\nu}-\partial_{\nu}A_{\mu}$, $\Lambda$ is the cosmological constant, and $\alpha$ is the Gauss-Bonnet parameter. A black hole solution consistent with the action in eq.(\ref{1.1}) reads \cite{DL}
\begin{equation}\label{1.3}
ds^2=-f(r)dt^2+\dfrac{dr^2}{f(r)}+r^2d\Omega_{D-2}^2
\end{equation}
where 
\begin{equation}\label{1.4}
\begin{split}
f(r)&=1+\dfrac{r^2}{2\alpha'}\left(1-\sqrt{1-\dfrac{4\alpha'}{l^2}+\dfrac{4\alpha'm}{r^{D-1}}-\dfrac{4\alpha'q^2}{r^{2D-4}}}\right)~.
\end{split}
\end{equation}
Here, $m$ is related with the AdM mass (M) of the black hole as 
\begin{equation}\label{Mass}
M=\dfrac{(D-2)\omega_{D-2}}{16\pi G}m ~~;~~\omega_{D-2}=\dfrac{2\pi^{(D-1)/2}}{\Gamma(D-1)/2} ~
\end{equation}
where $\omega_{D-2}$ is the volume of the unit sphere in $D-2$ dimensions. In eq.(\ref{1.4}), $\alpha'$ is a dimension dependent parameter which is related to the Gauss-Bonnet parameter $\alpha$ as $\alpha'=(D-3)(D-4)\alpha$.  In eq.(\ref{1.4}), the parameter $q$ is related to the total electric charge ($Q$) of the black hole  as 
\begin{equation}\label{charge}
Q=\sqrt{\dfrac{2(D-2)(D-3)}{G}}\dfrac{\omega_{D-2}q}{8\pi}~.
\end{equation}

\noindent Event horizon of the black hole is obtained from the relation $f(r_+)=0$ which relates the mass $M$ in terms of the horizon radius ($r_+$) as
\begin{equation}\label{1.6}
M=\dfrac{(D-2)\omega_{D-2}}{16\pi G}\left(\dfrac{r_+^{D-1}}{l^2}+r_+^{D-3}+\alpha'r_+^{D-5}+\dfrac{q^2}{r_+^{D-3}}\right)~.
\end{equation}
The Hawking temperature of the black hole from eq.(\ref{1.4}) reads
\begin{equation}\label{1.7}
\begin{split}
T&=\dfrac{1}{4\pi}\dfrac{\partial f}{\partial r}\bigg\vert_{r=r_+}\\&=\dfrac{(D-1)r_+^3}{4\pi l^2(r_+^2+2\alpha')}+\dfrac{(D-3)r_+}{4\pi(r_+^2+2\alpha')}+\dfrac{(D-5)\alpha'}{4\pi r_+(r_+^2+2\alpha')}\\
&-\dfrac{(D-3)q^2r_+^{7-2D}}{4\pi(r_+^2+2\alpha')}~.
\end{split}
\end{equation}
Gauss-Bonnet gravity theory becomes non-trivial only for spacetime dimensions $D\geq5$. Hence, we have carried out a detailed analysis of the phase structure of an AdS black hole in 5 dimensions (in section 4). For $D=5$, the Hawking temperature from eq.(\ref{1.7}) can be calculated as follows
\begin{equation}\label{teee}
T_5=\dfrac{r_+^3}{\pi l^2(r_+^2+2\alpha')}+\dfrac{r_+}{2\pi(r_+^2+2\alpha')}-\dfrac{q^2r_+^{-3}}{2\pi(r_+^2+2\alpha')}~.
\end{equation}
By invoking the first law of black hole thermodynamics ($dM=TdS$) and utilising eq.(s)(\ref{1.6},\ref{1.7}), we can calculate the entropy of the black hole as
\begin{equation}\label{ent}
S=\int_0^{r_+}T^{-1}\left(\dfrac{\partial M}{\partial r_+}\right)dr_+= \dfrac{\omega_{D-2}}{4G}r_+^{D-2}\left[1+\dfrac{(D-2)2\alpha'}{(D-4)r_+^2}\right]~.
\end{equation}
Having calculated all the thermodynamic variables for the black hole we are now in a position to find the ``\textit{mixed first law}" of black hole thermodynamics \cite{mann1} by utilising inputs from the conformal boundary field theory in the \textit{AdS}/CFT framework in the following section.
\section{The first law of thermodynamics and its modified form}
In this section we shall try to compute the first law of black hole thermodynamics in a mixed form. To do this we shall consider additional inputs from the boundary CFT theory and using them compute the effective volume and chemical potential. At first we shall compute the conjugate to the usual thermodynamic variables from the extended first law of black hole thermodynamics. In recent works it has been realized that one can induce a positive thermodynamical pressure by using a negative cosmological constant. For an underlying negative cosmological constant, one can rewrite the thermodynamics pressure of a black hole as follows \cite{kastor}
\begin{equation}\label{Pressure}
P=-\frac{\Lambda}{8\pi G}
\end{equation} 
where the cosmological constant has the form given by
\begin{equation}\label{Cosmological_Constant}
\Lambda=-\frac{(D-1)(D-2)}{2l^2}
\end{equation} 
with $l$ being the \textit{AdS} radius. The other thermodynamic variable and its conjugate are the Bekenstein-Hawking entropy and the Hawking temperature of the \textit{AdS} black hole in Gauss-Bonnet gravity. The Bekenstein-Hawking entropy in terms of $A$ and the Newton's gravitational constant reads (in natural units)  \cite{kastori} \footnote{Note that the $A$ in eq.(\ref{Area}) should not be confused with the area of the black hole.}
\begin{equation}\label{Entropy}
S=\frac{A}{4G}
\end{equation} 
where 
\begin{equation}\label{Area}
A=\omega_{D-2}r_+^{D-2}+\frac{2(D-2)\alpha'}{D-4}\omega_{D-2}r_+^{D-4}~.
\end{equation}
In eq.(\ref{Area}), the first term is equivalent to the area of the event horizon of the black hole in $D$-dimensional \textit{AdS} spacetime and second term arises due to the consideration of Gauss-Bonnet gravity. The Hawking temperature is the conjugate variable to the Bekenstein-Hawking entropy. In terms of the surface gravity $\kappa$  of the black hole, we can write the Hawking temperature as 
\begin{equation}\label{Hawking_Temperature}
T=\frac{\kappa}{2\pi}~.
\end{equation}
In this extended thermodynamics, using eq.(s)(\ref{Pressure},\ref{Entropy},\ref{Hawking_Temperature}), the most general form of the first law of black hole thermodynamics can be written as \cite{kastor}
\begin{equation}\label{Extended_first_law}
\begin{split}
\delta M=&T\delta S+V\delta P+\Phi \delta Q+\Omega\delta J\\=&\frac{\kappa}{8\pi G}\delta A-\frac{V}{8\pi G}\delta \Lambda+\Phi \delta Q+\Omega\delta J~.
\end{split}
\end{equation}
A direct interpretation of the above first law of black hole thermodynamics has some issues in the context of holography \cite{cv,DolanBP,Kastor2,Zhang1,Zhang2,DolanBP2,McCarthy,Jhonson2}. It is observed that for a single pressure-volume term in the first law of thermodynamics in the bulk, there are two terms in the first law of thermodynamics of the conformal field theory at the boundary, the central charge and its conjugate variable, and the thermodynamical pressure and its conjugate variable. In order to counter this issue one uses the \textit{AdS}/CFT dictionary. Via the \textit{AdS}/CFT correspondence one can relate the central charge corresponding to the boundary field theory with the \textit{AdS} radius ($l$) and Newton's gravitational constant ($G$) in the bulk theory. The form of the central charge $C$ is given as follows \cite{karch}
\begin{equation}\label{Central_Charge}
C=k\frac{l^{D-2}}{16\pi G}~.
\end{equation}  
In eq.(\ref{Central_Charge}), the $k$ factor is fixed by the details of the system at the boundary. From the above equation it can be seen that by varying one of the bulk variables one cannot keep the central charge fixed. Therefore, to keep $C$ fixed, one needs to vary not only the Newtonian gravitational constant $G$ but also the \textit{AdS} radius $l$. In our analysis all the thermodynamic variables being varied are all dimensionful variables. Being a dimensionful parameter, we shall extend our analysis by including the Gauss-Bonnet parameter as a thermodynamic variable in the mixed first law of thermodynamics as well. To begin our analysis we write the mass of the black hole as 
\begin{equation}\label{BH_Mass}
M=M(A,G,J,Q,\Lambda,\alpha)
\end{equation}
where we have considered the mass of the black hole as a function of $A$, Newton's gravitational constant ($G$), angular momentum ($J$), total charge of the black hole ($Q$), the cosmological constant ($\Lambda$) and the Gauss-Bonnet parameter ($\alpha$). Taking variation of both sides of eq.(\ref{BH_Mass}), we obtain
\begin{equation}\label{Mixed_1}
\delta M=\frac{\partial M}{\partial A}\delta A+\frac{\partial M}{\partial G}\delta G+\frac{\partial M}{\partial J}\delta J+\frac{\partial M}{\partial Q}\delta Q+\frac{\partial M}{\partial \Lambda}\delta \Lambda+\frac{\partial M}{\partial \alpha}\delta \alpha.
\end{equation} 
For a fixed value of the Gauss-Bonnet parameter ($\alpha$) and the Newton's gravitational constant ($G$), the above form of the first law of thermodynamics reduces to the known extended first law of thermodynamics in the bulk theory. Hence, by comparing eq.(\ref{Mixed_1}) with eq.(\ref{Extended_first_law}) in this scenario, we can obtain the following relations
\begin{align}
\frac{\partial M}{\partial A}=\frac{\kappa}{8\pi G}~,~~\frac{\partial M}{\partial J}=\Omega~,~~\frac{\partial M}{\partial Q}=\Phi~,~~\frac{\partial M}{\partial \Lambda}=-\frac{V}{8\pi G}~.\label{Conjugate_Variables}
\end{align} 
We define the conjugate variables to the Newton's gravitational constant $G$ and the Gauss-Bonnet parameter ($\alpha$) as 
\begin{equation}\label{New_Conjugate_Variables}
\frac{\partial M}{\partial G}\equiv-\frac{\zeta}{G}~,~~\frac{\partial M}{\partial \alpha}\equiv\mathcal{A}~. 
\end{equation}
Using the analytical forms of the conjugate variables in eq.(s)(\ref{Conjugate_Variables},\ref{New_Conjugate_Variables}), we can rewrite eq.(\ref{Mixed_1}) as follows
\begin{equation}\label{Mixed_2_with_Conjugate}
\delta M=\frac{\kappa}{8\pi G}\delta A+\Omega\delta J+\Phi\delta Q-\frac{V}{8\pi G}\delta \Lambda-\zeta\frac{\delta G}{G}+\mathcal{A}\delta \alpha~.
\end{equation}
In order to investigate the analytical form of the unknown parameter $\zeta$, we shall consider a modified mass term as follows
\begin{equation}\label{Modified_Mass}
GM=\mathcal{M}=\mathcal{M}(A,GJ,\sqrt{G}Q,\Lambda,\alpha)~.
\end{equation}
It is very important to observe that for the modified mass term ($\mathcal{M}$), the Newtonian gravitational constant no longer acts as an independent variable rather it gets coupled to the total charge and the angular momentum of the black hole. Again taking variation of eq.(\ref{Modified_Mass}), and some algebra, we obtain the following result
\begin{equation}\label{Mixed_Modified_2}
\begin{split}
&G\delta M+M\delta G=\frac{\partial \mathcal{M}}{\partial A}\delta A+J\frac{\partial \mathcal{M}}{\partial (GJ)}\delta G+G\frac{\partial \mathcal{M}}{\partial (GJ)}\delta J+\\&\sqrt{G}\frac{\partial \mathcal{M}}{\partial (\sqrt{G}Q)}\delta Q+\frac{Q}{2\sqrt{G}}\frac{\partial \mathcal{M}}{\partial (\sqrt{G}Q)}\delta G+\frac{\partial \mathcal M}{\partial \Lambda}\delta \Lambda+\frac{\partial \mathcal{M}}{\partial \alpha}\delta \alpha~. 
\end{split}
\end{equation}
Rearranging the above equation, we can obtain the variation of the unmodified mass of the black hole ($M$) as follows
\begin{equation}\label{Rearranged_Modified_Mixed_2}
\begin{split}
\delta M&=\frac{1}{G}\frac{\partial \mathcal{M}}{\partial A}\delta A+\frac{\partial \mathcal{M}}{\partial (GJ)}\delta J+\frac{1}{\sqrt{G}}\frac{\partial \mathcal{M}}{\partial (\sqrt{G}Q)}\delta Q+\frac{\partial \mathcal{M}}{\partial \Lambda}\frac{\delta \Lambda}{G}\\&+\frac{\partial \mathcal{M}}{\partial \alpha}\frac{\delta \alpha}{G}+\left[-\frac{M}{G}+\frac{Q}{2G^{\frac{3}{2}}}\frac{\partial \mathcal{M}}{\partial (\sqrt{G}Q)}+\frac{J}{G}\frac{\partial \mathcal{M}}{\partial (GJ)}\right]\delta G~.
\end{split}
\end{equation}
Now eq.(\ref{Mixed_2_with_Conjugate}) along with eq.(\ref{Rearranged_Modified_Mixed_2}) denotes the same first law of thermodynamics. Hence, by comparing eq.(\ref{Rearranged_Modified_Mixed_2}) along with eq.(\ref{Mixed_2_with_Conjugate}), we obtain the following  results
\begin{equation}\label{Modified_Conjugates}
\begin{split}
\frac{\partial \mathcal{M}}{\partial A}&=\frac{\kappa}{8\pi}~,~\frac{\partial \mathcal{M}}{\partial (GJ)}=\Omega~,~\frac{\partial \mathcal{M}}{\partial (\sqrt{G}Q)}=\sqrt{G}\Phi~,\\\frac{\partial \mathcal{M}}{\partial \Lambda}&=-\frac{V}{8\pi}~,~\frac{1}{G}\frac{\partial \mathcal{M}}{\partial \alpha}= \mathcal{A}~.
\end{split}
\end{equation}
We also obtain the analytical form of the unknown parameter $\zeta$ as
\begin{equation}\label{zeta_form}
\zeta=M-\frac{Q}{2\sqrt{G}}\frac{\partial \mathcal{M}}{\partial (\sqrt{G}Q)}-J\frac{\partial \mathcal{M}}{\partial (GJ)}~.
\end{equation}
Using the values of $\frac{\partial \mathcal{M}}{\partial (\sqrt{G}Q)}$ and $\frac{\partial\mathcal{M}}{\partial (GJ)}$ from eq.(\ref{Modified_Conjugates}), we can rewrite eq.(\ref{zeta_form}) as follows
\begin{equation}\label{zeta_final}
\zeta=M-\frac{Q\Phi}{2}-\Omega J~.
\end{equation}
To proceed further we take the variation of eq.(\ref{Central_Charge}) and divide it by the form of the central charge to get
\begin{equation}\label{C_Variation}
\frac{\delta C}{C}=-\frac{\delta G}{G}+(D-2)\frac{\delta l}{l}~.
\end{equation}
Using the forms of the thermodynamic pressure ($P$) and the cosmological constant ($\Lambda$), we obtain the form of $\frac{\delta l}{l}$ as follows
\begin{equation}\label{dl_by_l}
\frac{\delta l}{l}=-\frac{\delta G}{G}-\frac{\delta P}{2P}~.
\end{equation} 
Using the above equation in eq.(\ref{C_Variation}), we obtain the variation in $G$ in terms of the variation of the central charge and the thermodynamic pressure as follows
\begin{equation}\label{dG_by_G}
\frac{\delta G}{G}=-\frac{2}{D}\frac{\delta C}{C}-\frac{D-2}{D}\frac{\delta P}{P}~.
\end{equation}
Now substituting eq.(s)(\ref{Modified_Conjugates},\ref{zeta_final},\ref{dG_by_G}) in eq.(\ref{Rearranged_Modified_Mixed_2}), we obtain the form of the modified first law as follows
\begin{equation}\label{Modified_Pre_final}
\begin{split}
\delta M=&\frac{\kappa}{8\pi G}\delta A+\Omega \delta J+\Phi \delta Q-\frac{V}{8\pi G}\delta \Lambda+\mathcal{A}\delta \alpha+\frac{2\zeta}{DC}\delta C\\+&\frac{D-2}{D}\zeta\frac{\delta P}{P}\\
=&T\delta S+\Omega\delta J+\Phi\delta Q+\left[\frac{2\zeta}{DC}-\frac{2(TS+PV)}{DC}\right]\delta C\\+&\mathcal{A}\delta\alpha+\left[V+\frac{D-2}{DP}\zeta-\frac{D-2}{DP}(TS+PV)\right]\delta P
\end{split}
\end{equation} 
where we have made use of eq.(s)(\ref{Pressure},\ref{Entropy},\ref{Hawking_Temperature},\ref{dG_by_G}). We can rewrite eq.(\ref{Modified_Pre_final}) as follows
\begin{equation}\label{Mixed_final}
\delta M=T\delta S+\Omega\delta J+\Phi\delta Q+\mathcal{A}\delta\alpha+V_\mathcal{C}\delta P+\mu_\mathcal{C}\delta C
\end{equation}
where $V_\mathcal{C}$ and $\mu_\mathcal{C}$ gives the effective thermodynamic volume and chemical potential given by
\begin{align}
V_\mathcal{C}&=V+\frac{D-2}{DP}\zeta-\frac{D-2}{DP}(TS+PV)~,\label{Effective_Volume}\\
\mu_\mathcal{C}&=\frac{2\zeta}{DC}-\frac{2(TS+PV)}{DC}~.
\end{align}
We can see that the form of the first law of black hole thermodynamics obtained in eq.(\ref{Mixed_final}) contains contributions from the bulk and boundary variables. Therefore, we call it a mixed form of the first law of thermodynamics for a Gauss-Bonnet \textit{AdS} black hole.
\subsection{Extended black hole thermodynamics and the Smarr relation}
Using the extended first law of black thermodynamics and the Smarr relation, we would now like to obtain the analytical form of $V$ in terms of the mass $M$ of the black hole, angular momentum $J$, total charge $Q$ of the black hole and the Gauss-Bonnet parameter $\alpha$. The extended first law of black hole thermodynamics for a Gauss-Bonnet \textit{AdS} black hole is given by
\begin{equation}\label{1.28}
\delta M=T\delta S+\Phi \delta Q+\Omega\delta J+V\delta P+\mathcal{A}\delta \alpha~.
\end{equation} 
It is pretty straightforward to infer the following relations
\begin{align}\label{1.29}
T=\frac{\delta M}{\delta S}~,~\Phi=\frac{\delta M}{\delta Q}~,~V=\frac{\delta M}{\delta P}~,~\Omega=\frac{\delta M}{\delta J}~,~\mathcal{A}=\frac{\delta M}{\delta \alpha}~.
\end{align}
Here we have considered the black hole mass to be a function of the entropy ($S$), angular momentum ($J$), thermodynamic pressure ($P$), charge ($Q$) and the Gauss-Bonnet parameter ($\alpha$) ($M=M(S,J,P,Q,\alpha)$). $M,S,J,P,Q,\alpha$ have the following dimensions (in terms of $L$) in generalized $D$-dimensions (with $G$ being considered as a dimensionless quantity)
\begin{equation}\label{1.30}
\begin{split}
[M]&=L^{D-3},~[S]=L^{D-2},~[J]=L^{D-2}~,\\
[P]&=L^{-2}~,~[Q]=L^{D-3},~[\alpha]=L^{2}~.
\end{split}
\end{equation}
\noindent Now making use of Euler's theorem of quasi-homogeneous functions, we get the following relation
\begin{equation}\label{1.31}
\begin{split}
(D-3)M&=(D-2)S\frac{\delta M}{\delta S}+2\alpha\frac{\delta M}{\delta \alpha}-2P\frac{\delta M}{\delta P}\\&+(D-3)Q\frac{\delta M}{\delta Q}+(D-2)J\frac{\delta M}{\delta J}~.
\end{split}
\end{equation}
\noindent Substituting the forms of the conjugate variables to $S,Q,P,$ $J,\alpha$ in eq.(\ref{1.31}), we get the desired Smarr relation as 
\begin{equation}\label{1.32}
(D-3)M=(D-2)TS+2\mathcal{A} \alpha-2PV+(D-3)\Phi Q+(D-2)\Omega J~.
\end{equation}
Rearranging the above equation, one can obtain the volume of the black hole in terms of $T,~S,~P,~\mathcal{A},~\alpha,~Q,~M,~\Omega,$ and $J$ as follows
\begin{equation}\label{1.33}
V=\frac{D-2}{2P}TS+\frac{\mathcal{A}\alpha}{P}+\frac{D-3}{2P}\Phi Q-\frac{D-3}{2P}M+\frac{D-2}{2P}\Omega J.
\end{equation}
Replacing the form of $V$ from eq.(\ref{1.33}) in eq.(\ref{Effective_Volume}), we get
\begin{equation}\label{1.34}
V_\mathcal{C}=\frac{2M+4\Gamma \alpha+(D-4)Q\Phi}{2DP}~.
\end{equation}  
 $V_\mathcal{C}$ is now the new thermodynamical volume for the Gauss-Bonnet $AdS$ black hole which depends on the Gauss-Bonnet parameter $\alpha$ which enters through the expression for $V$ obtained from the Smarr formula. This new definition of the thermodynamic volume on account of modification to the gravity theory should have effects on the phase structure of the black holes. Black holes in extended phase space in Gauss-Bonnet gravity are known to have Van der Waals gas like behaviour \cite{rongPV} and show critical behaviour. Our next goal is to understand the effects on the critical behaviour of the black hole in this new setup of mixed first law and the possible phase transition structure.
 \section{Phase transition structure}
\begin{figure}[ht!]
\centering
\includegraphics[scale=0.36]{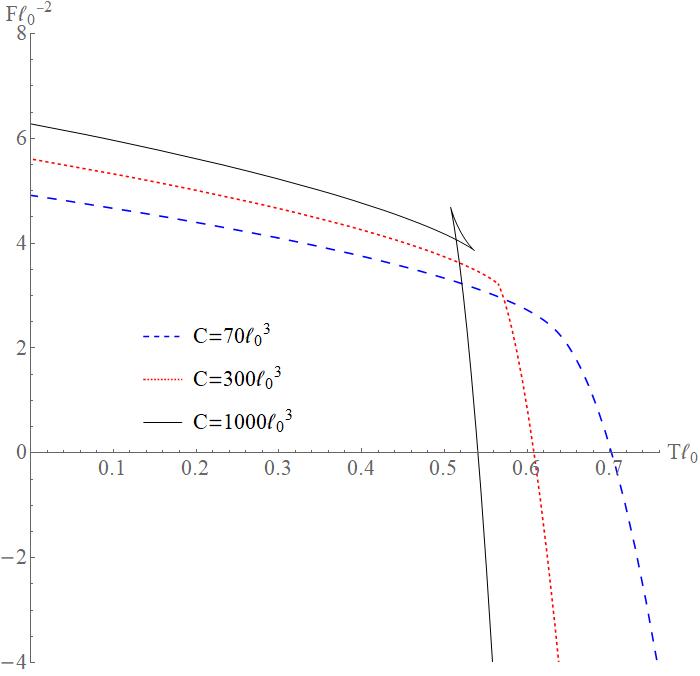}
\caption{Free energy versus temperature for different values of the central charge: $Q=1.0\mathcal{l}_0^2$; $k=16\pi$; $P\mathcal{l}_0^{2}=15$; $\alpha=0.001\mathcal{l}_0^2$; $D=5$} 
\label{f11}
\end{figure}

\begin{figure}[ht!]
\centering
\includegraphics[scale=0.36]{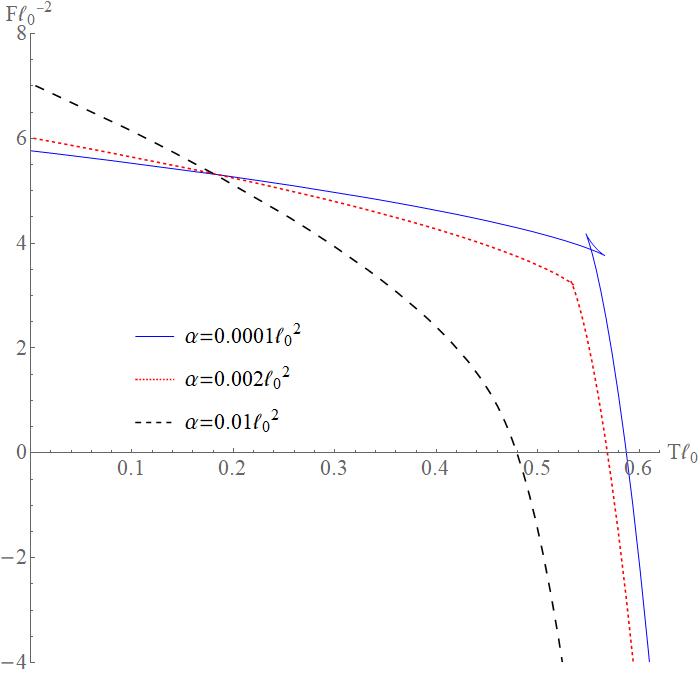}
\caption{Free energy versus temperature for different values of the Gauss-Bonnet parameter: $Q=1.0\mathcal{l}_0^2$; $k=16\pi$; $P\mathcal{l}_0^{2}=15$; $C=500\mathcal{l}_0^3$; $D=5$} 
\label{f22}
\end{figure}

In this section, we will try to investigate the phase transition structure of the black hole. We shall be interested in the dependence of phase transition on parameter $\alpha$. The free energy of the black hole can be computed using eq.(s)(\ref{1.7},\ref{Cosmological_Constant}) along with the entropy formula derived in eq.(\ref{ent}) as follows
\begin{equation}
\begin{split}\label{1.38}
F&=M-TS\\
&=\dfrac{(D-2)\omega_{D-2}}{16\pi G}\left(\dfrac{r_+^{D-1}}{l^2}+r_+^{D-3}+\alpha'r_+^{D-5}+\dfrac{q^2}{r_+^{D-3}}\right)\\&-\dfrac{\omega_{D-2}r^{D-3}}{16\pi G(r_+^2+2\alpha')}\left(\dfrac{(D-4)r_+^2+2(D-2)\alpha'}{(D-4)r_+^2}\right)\\&\left(\dfrac{(D-1)r_+^4}{l^2}+(D-3)r_+^2+(D-5)\alpha'-\dfrac{(D-3)q^2}{r_+^{2D-8}}\right)~.
\end{split}
\end{equation}
Specifically, for a five dimensional \textit{AdS} spacetime the form of free energy becomes
\begin{eqnarray}
F_5=\dfrac{3\pi}{8G}\left(\dfrac{r_+^4}{l^2}+r_+^2+\alpha'+\dfrac{q^2}{r_+^2}\right)-\dfrac{\pi (r_+^2+6\alpha')}{4G(r_+^2+2\alpha')}
\nonumber\\
\left(\dfrac{2r_+^4}{l^2}+r_+^2-\dfrac{q^2}{r_+^2}\right)
\end{eqnarray}
Here, we can express the free energy as a function of $T,$ $P,$ $Q,$ $C,$ and $\alpha$. In Fig. (\ref{f11}), we have plotted the free energy ($F$) with respect to the temperature ($T$) for different values of the central charge ($C$) keeping $Q$, $k$, $\alpha$, and $P$ fixed. From Fig.(\ref{1.1}), we observe that above a certain critical value of the central charge, the free energy curve attains a swallow-tail like structure with respect to change in the temperature $T$. The behaviour is similar to black holes in Einstein's gravity as discussed in \cite{mann1, mann2}. In Fig. (\ref{1.1}), we have used  $Q=1.0\mathcal{l}_0^2$, $k=16\pi$, $P\mathcal{l}_0^{2}=15$, $\alpha=0.001\mathcal{l}_0^2$, and $D=5$ for an arbitrary length scale $\mathcal{l}_0$. One crucial difference in our analysis is that  the critical central charge value is not universal any more. This breaking of universality is attributed to the Gauss-Bonnet term and higher dimensions as discussed in the following section. Hence, from Fig.(\ref{f11}) we observe that for $AdS$ black holes in Gauss-Bonnet gravity, there exists a critical value of the central charge for fixed values of pressure and the Gauss-Bonnet parameter above which there exists a small to large black hole phase transition.
\noindent For the next part of our analysis, we have concentrated on the dependence of the critical value of the central charge on the Gauss-Bonnet parameter $\alpha$. We have plotted free energy with respect to the temperature for different values of the Gauss-Bonnet parameter $\alpha$ for a fixed value of the central charge. From Fig.(\ref{f22}), we observe that the presence of the parameter $\alpha$ shifts the critical point. The values of the parameters used in Fig.(\ref{f22}) are $Q=1.0\mathcal{l}_0^2,~ k=16\pi,~ P\mathcal{l}_0^2=15,~C=500\mathcal{l}_0^3,$ and $D=5$ for an arbitrary length scale $\mathcal{l}_0$. In Fig.(\ref{f22}), decrease in the Gauss-Bonnet parameter results in the increase of the critical temperature ($T_c$) values. Hence, the Gauss-Bonnet parameter has a significant effect near the critical point. 
\begin{figure}[ht!]
\centering
\includegraphics[scale=0.36]{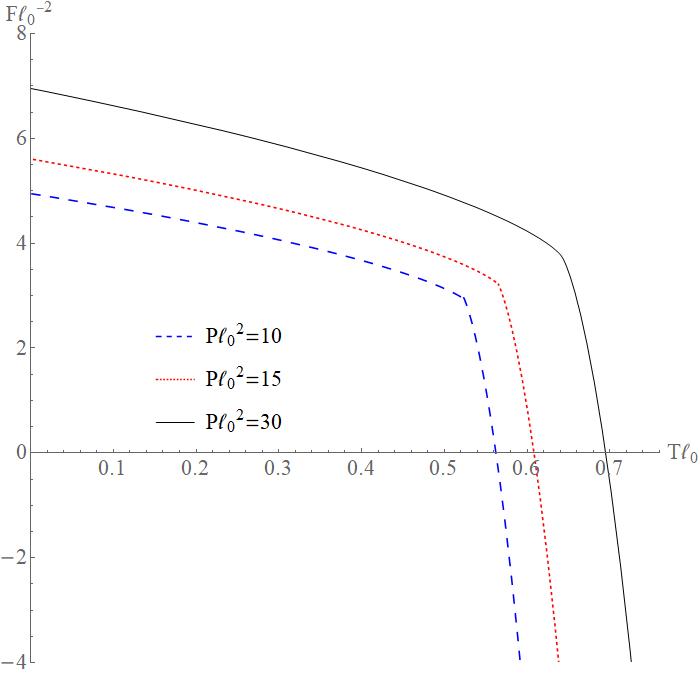}
\caption{Free energy versus temperature for different values of pressure: $Q=1.0\mathcal{l}_0^2$; $k=16\pi$; $\alpha=0.001\mathcal{l}_0^{2}$; $C=300\mathcal{l}_0^3$; $D=5$} 
\label{f23}
\end{figure}



\noindent In Fig.(\ref{f23}), we have plotted the free energy of the black hole with respect to the temperature for different values of the thermodynamic pressure. We have used $Q=1.0\mathcal{l}_0^2$, $k=16\pi$, $\alpha=0.001\mathcal{l}_0^2$, and $C=300\mathcal{l}_0^3$ as the values of the other parameters. It is very important to observe that with the increase in the thermodynamic pressure the phase transition point shifts more to the higher temperature region. Therefore, to obtain a phase transition behaviour in lower temperature region, one needs to lower the thermodynamic pressure as well which is proportional to the negative value of the cosmological constant (eq.(\ref{Pressure})).  This pressure dependence reaffirms the Van der Waals gas like behaviour. 
\begin{figure}[ht!]
\centering
\includegraphics[scale=0.36]{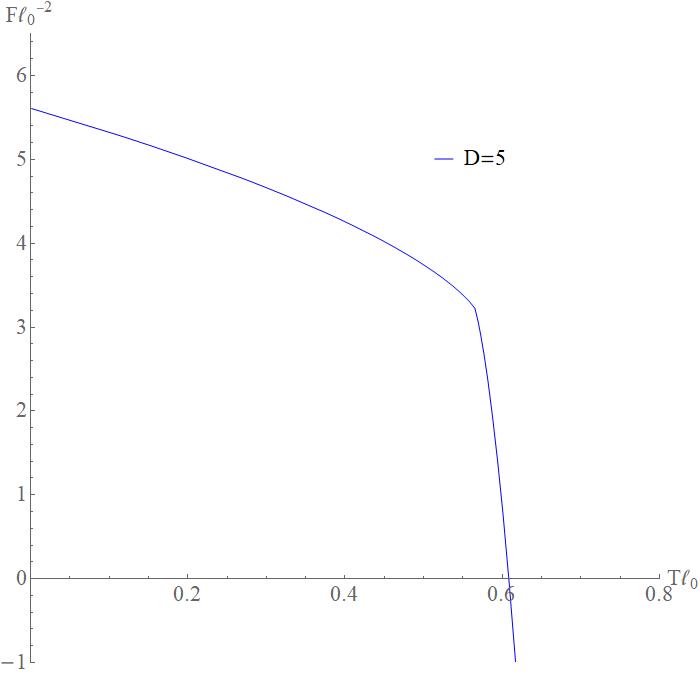}
\caption{Free energy versus temperature in five spacetime dimensions: $Q=1.0\mathcal{l}_0^2$; $k=16\pi$; $\alpha=0.001\mathcal{l}_0^{2}$; $C=300\mathcal{l}_0^3$; $D=5$} 
\label{f24a}
\end{figure}
\begin{figure}[ht!]
\centering
\includegraphics[scale=0.36]{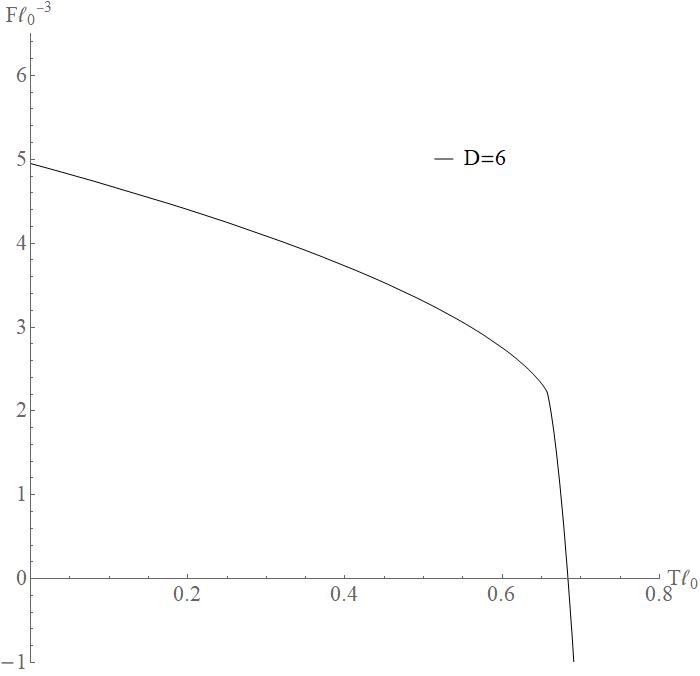}
\caption{Free energy versus temperature in six spacetime dimensions: $Q=1.0\mathcal{l}_0^3$; $k=16\pi$; $\alpha=0.001\mathcal{l}_0^{2}$; $C=450\mathcal{l}_0^4$; $D=6$} 
\label{f24b}
\end{figure}
\begin{figure}[ht!]
\centering
\includegraphics[scale=0.36]{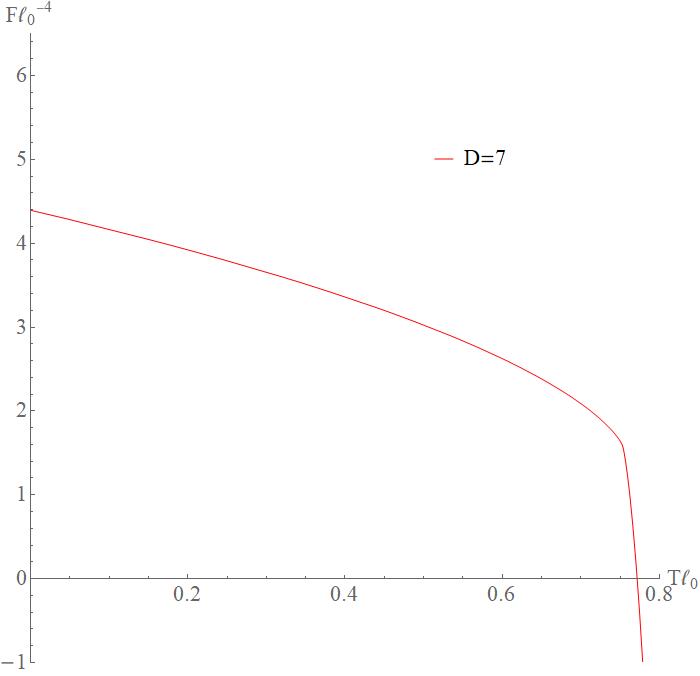}
\caption{Free energy versus temperature in seven spacetime dimensions: $Q=1.0\mathcal{l}_0^4$; $k=16\pi$; $\alpha=0.001\mathcal{l}_0^{2}$; $C=600\mathcal{l}_0^5$; $D=7$} 
\label{f24c}
\end{figure}
\noindent Till now we have exploited the phase transition behaviour in five dimensional spacetime only. Since, our analysis is done in general $D$-spacetime dimensions, we are in a position to analyse the phase transition behaviour for different values of $D$. Fig.(s)(\ref{f24a},\ref{f24b},\ref{f24c}) shows the behaviour of free energy with the Hawking temperature of the black hole for $D=5,~6,~7$. In these three figures we have plotted the free energy behaviour with respect to the temperature around the critical value of the central charge in different dimension. It is very important to observe that the critical behaviour is attained at higher values of the temperature when in higher dimensions. 

\noindent It is important to note that our finding here is very similar to what we obtained in our earlier work \cite{nss}. We also observe some differences within the results as well. There we observed that for a Born-Infeld $AdS$ black hole in four spacetime dimensions, a no-black hole region below a critical temperature with a sufficiently small value of the Born-Infeld parameter in case of the free energy versus temperature plot. Here, in the case of Gauss-Bonnet gravity in five spacetime dimensions, we observe no such region even with the increase in the Gauss-Bonnet parameter $\alpha$.

\noindent We would like to make another comment. The phase transition structure that we observe here is very similar to that obtained in \cite{rongPV}. However, the structure found here differs from that in \cite{rongPV} due to the involvement of the central charge in the first law of black hole thermodynamics, also referred to as the mixed first law. The phasespace is even more extended now, the critical point of phase transition involves an additional parameter $C_c$ (critical central charge) along with pressure, volume and temperature. The qualitative behaviour of the black hole phase transition is still the same as the Van der Waals fluid. The role of the central charge is to shift the phase transition point which in principle can be a detectable signature.
\section{Breaking of universal nature of the central charge}
\noindent We shall calculate the critical  value of the central charge for $D=5$ spacetime dimensions. In order to find this value, we will use the following two equations\cite{mann1}
\begin{align}
\dfrac{\partial T_5}{\partial r_+}&=0~,\label{1.39}\\
\dfrac{\partial^2T_5}{\partial r_+^2}&=0\label{1.40}~.
\end{align}
Using eq.(\ref{teee}) in the above two equations, we get
\begin{equation}\label{t51}
2r^8_{+(c)}+(12\alpha' -l_c^2)r^6_{+(c)}+2l^2_c\alpha'  r^4_{+(c)}+5l_c^2q^2r^2_{+(c)}+6l^2_cq^2\alpha' =0
\end{equation}
and
\begin{eqnarray}\label{t52}
(l_c^2-4\alpha')r^8_{+(c)}+(24\alpha'^2-6\alpha'l^2_c)r^6_{+(c)}-15q^2l_c^2r^4_{+(c)}
\nonumber\\
-34\alpha'q^2l_c^2r^2_{+(c)}-24\alpha'^2q^2l_c^2=0~.
\end{eqnarray}
Here, $c$ in the subscript of $r_{+(c)}$ and $l_c$ is denoting the critical value of these parameters. \
\noindent These two equations are difficult to solve exactly, however, we are interested in the qualitative behaviour of the solution in order to understand the universal behaviour of central charge and its dependence on the parameter $\alpha$. Hence, we solve these to first order in $\alpha'$ perturbatively. We take a solution of the form 
\begin{equation}\label{1.41b}
r_{+(c)}\cong r_{+(c)}^{(0)}+\alpha' r_{+(c)}^{(1)}~,~~l_c\cong  l_c^{(0)}+\alpha' l_c^{(1)}.
\end{equation}
Putting these back in eq.(s)(\ref{t51}, \ref{t52}), we obtain the forms of $r_{+(c)}^{(0)}$ and $l_c^{(0)}$ to be
\begin{equation}\label{1.41c}
r_{+(c)}^{(0)}=15^{1/4}\sqrt{q}~,~~l_c^{(0)}=3^{3/4}5^{1/4}\sqrt{q}~.
\end{equation}
Using the values of $r^{(0)}_{+(c)}$ and $l_c^{(0)}$ from eq.(\ref{1.41c}) in eq.(s)(\ref{1.39},\ref{1.40}) and solving it perturbatively upto $\mathcal{O}(\alpha)$, one can obtain the forms of $r_{+(c)}^{(1)}$ and $l_{c}^{(1)}$ as follows
\begin{equation}\label{1.42a}
r_{+(c)}^{(1)}=\frac{3^{\frac{3}{4}}4}{5^{\frac{5}{4}}\sqrt{q}}~,~~l_{c}^{(1)}=\frac{3^{\frac{1}{4}}24}{5^{\frac{5}{4}}\sqrt{q}}~.
\end{equation}
The value of the parameter $q$ in terms of net electric charge on the black hole is given by eq.(\ref{charge}) which for a five dimensional \textit{AdS} spacetime reduces to the following form
\begin{equation}\label{q55}
q=2\frac{\sqrt{G}Q}{\sqrt{3}\pi}~.
\end{equation}
Using the form of $q$ from eq.(\ref{q55}) and the parameter $\alpha'$ in terms of $\alpha$ in $D=5$, we obtain the critical values of $r_+$ and $l$ as follows
\begin{align}
r_{+(c)}=&5^{\frac{1}{4}}\sqrt{\frac{2}{\pi}}\sqrt{Q}G^{\frac{1}{4}}+\frac{12\alpha\sqrt{2\pi}}{5^{\frac{5}{4}}\sqrt{Q}G^{\frac{1}{4}}}\label{Critr}~,\\
l_c=&5^{\frac{1}{4}}\sqrt{\frac{6}{\pi}}\sqrt{Q}G^{\frac{1}{4}}+\frac{24\sqrt{6\pi}\alpha}{5^{\frac{5}{4}}\sqrt{Q}G^{\frac{1}{4}}}~.\label{Critl}
\end{align}
Putting the values of $r_{+(c)}$ and $l_c$ in eq.(\ref{teee}), we can obtain the critical value of the temperature upto $\mathcal{O}(\alpha)$.
However, the most important result is the value of the critical central charge which can be obtained using the form of $l_c$ from eq.(\ref{Critl}) in eq.(\ref{Central_Charge}) for $D=5$. Upto $\mathcal{O}(\alpha)$, the value of the central critical charge becomes 
\begin{equation}\label{1.45}
\begin{split}
C_c&=k\dfrac{l_c^3}{16\pi G}\\
\implies C_c&\cong k\dfrac{l_c^{(0)^3}+3\alpha' l_c^{(1)}}{16\pi G}=k\frac{3^{\frac{3}{2}}5^{\frac{3}{4}}Q^{\frac{3}{2}}}{(2\pi)^{\frac{5}{2}}G^{\frac{1}{4}}}+\alpha k\frac{27\sqrt{6} \sqrt{Q} }{5^{\frac{3}{4}}G^{\frac{3}{4}}\pi^{\frac{3}{2}}}~.
\end{split}
\end{equation}
It is quite clear from eq.(\ref{1.45}) that unlike the expression of critical central charge in \cite{mann1} which just depends on charge $Q$, this has dependence on $G$ and the Gauss-Bonnet parameter $\alpha$. Hence, the universal nature of the central charge at critical point breaks. This is an important finding in this paper.

\noindent An important remark about the critical central charge in higher dimensions in general relativity is made in \cite{mann1}. It was speculated that the universality of central critical charge breaks in higher dimensions. Here, we are in a  position to check this fact. In the limit $\alpha\rightarrow 0$, $C_c$ becomes (in $D=5$ spacetime dimensions)
\begin{equation}
C_c=k\dfrac{l_c^{(0)^3}}{16\pi G}~.
\end{equation}
Using eq.(s)(\ref{1.41c},\ref{q55}) in the above equation, we get
\begin{equation}\label{c555}
C_c=k\frac{3^{\frac{3}{2}}5^{\frac{3}{4}}Q^{\frac{3}{2}}}{(2\pi)^{\frac{5}{2}}G^{\frac{1}{4}}}~.
\end{equation}
 It  gives the form of the critical charge in terms of $Q$ and $G$ which manifests the speculation made in \cite{mann1} about breaking of the central charge criticality in higher dimensions due the presence of $G$ besides $Q$.\
 

\noindent Thus, the result in eq.(\ref{1.45}) manifests that central charge criticality breaks in Gauss-Bonnet gravity. Also, in the limit ($\alpha\rightarrow 0$) we get the behaviour of central charge near the critical point in Einstein's gravity in spacetime dimensions five. Universal nature of the critical central charge is seen to break here as well (eq.\ref{c555}). In an Appendix, we investigate whether the universality of the critical central charge remains in dimensions greater than four or not.

\noindent If we compare this result with the Born-Infeld case \cite{nss}, we observe that the universal nature of the central charge broke there only due to the inclusion of the Born-Infeld parameter. Here, in case of five dimensional Gauss-Bonnet gravity even without the Gauss-Bonnet parameter the central charge has no universal behaviour. Hence, the universal nature of the critical value of the central charge seems to be a special feature of Einstein-Hilbert action with Maxwell-fields and cosmological constant in $3+1$-spacetime dimensions as it has been shown to break in higher dimensions and also with non-linear gauge fields and higher curvature terms. 
 

\section{Conclusion}
In this work we have investigated the thermodynamics of a Gauss-Bonnet \textit{AdS} black holes in general $D$-dimensions. In our analysis we have varied the Newton's gravitational constant, $AdS$ radius, and the Gauss-Bonnet parameter. Our central analysis involves the derivation of the mixed first law of black hole thermodynamics involving thermodynamic variables from both the boundary and the bulk. As an additional input we have considered the Gauss-Bonnet parameter as a thermodynamic variable as well. Being a dimensionful variable we have considered the Gauss-Bonnet parameter while writing the modified Smarr relation and as a result  the Gauss-Bonnet parameter appeared in the analytical form of the modified thermodynamical variables $V_\mathcal{C}$ and $\mu_\mathcal{C}$. Due to this modification the central charge has a direct dependence on the Gauss-Bonnet parameter. It is very important to observe that in $D=5$ spacetime dimensions the critical value of the central charge is no more universal even for a vanishing value of the Gauss-Bonnet parameter. Hence, we can conclude that the universality of the critical value of the central charge is an unique property of the $D=4$ spacetime dimensions \cite{mann1}. This is one of the most important results in our paper. Next we have plotted the free energy vs temperature for a fixed value of the thermodynamical pressure. We observe that due to the inclusion of the Gauss-Bonnet parameter, the phase transition structure undergoes crucial change. Free energy analysis is done for $D=5$ since the Gauss-Bonnet gravity has non-trivial contribution only in dimensions $D\geq 5$. It implies that the inclusion of the Gauss-Bonnet term results in some significant change in the overall free energy behaviour of the black hole with respect to the change in the Hawking temperature. 


\section*{Appendix: Breaking of universal nature of critical central charge in dimensions greater than four}
\noindent The Hawking temperature in general $D$-dimensional Einstein gravity can be obtained from eq.(\ref{1.7}) in the limit $\alpha\rightarrow 0$, and reads
\begin{eqnarray}
T=\dfrac{1}{4\pi}\left(\dfrac{(D-1)r_+}{l^2}+\dfrac{(D-3)}{r_+}-\dfrac{(D-3)q^2}{r_+^{2D-5}}\right)~.
\end{eqnarray}  

\noindent The critical point of the phase transition corresponds to the point given by
\begin{eqnarray}
\dfrac{\partial T}{\partial r_+}=\dfrac{\partial^2T}{\partial r_+^2}=0~.
\end{eqnarray}
Solving the above equations, we get
\begin{equation}
r_c=\left[(2D-5)(D-2)q^2\right]^{1/(2D-6)}
\end{equation}
and
\begin{equation}
l_c=h(D)q^{1/(D-3)}~
\end{equation}
where 
\begin{equation}
h(D)=\dfrac{(D-1)^{1/2}(2D-5)^{1/2(D-3)}(D-2)^{(D-2)/2(D-3)}}{(D-3)}~.
\end{equation}
 Substituting the value of $l_c$ in eq.(\ref{Central_Charge}) and using eq.(\ref{charge}), we get the critical central charge of the form
\begin{equation}\label{Central_Charge_Modified}
C_c=\kappa\dfrac{g(D)G^{(D-2)/2(D-3)}Q^{(D-2)/(D-3)}}{16\pi G}~.
\end{equation}
Here, the dimension dependent factor is given by 
\begin{equation}\label{gD}
g(D)=h^{D-2}\left(\dfrac{8\pi}{\omega_{D-2}\sqrt{2(D-2)(D-3)}}\right)^{\frac{(D-2)}{(D-3)}}~.
\end{equation}
Eq.(\ref{Central_Charge_Modified}) is $G$ independent only for dimensions $D=4$. Hence, the universal nature of the critical central charge is a feature of $4D$-Einstein gravity and breaks for dimensions greater than four.


\begin{thebibliography}{8}
\bibitem{mann1}W. Cong, D. Kubiznak and R. B. Mann; \href{https://doi.org/10.1103/PhysRevLett.127.091301}{Phys. Rev. Lett. 127 (2021) 091301}.

\bibitem{kastori} D. Kastor, S. Ray and J. Traschen; \href{https://doi.org/10.1088/0264-9381/27/23/235014}{Class. Quantum Grav. 27 (2010) 235014}.

\bibitem{uni1} W. Israel, \href{https://doi.org/10.1103/PhysRev.164.1776}{Phys. Rev. 164 (1967) 1776}.

\bibitem{uni11} W. Israel, \href{https://doi.org/10.1007/BF01645859}{Commun. Math. Phys. 8 (1968) 245}.

\bibitem{uni2}H. Muller Zum Hagen, D. C. Robinson, and H. J. Seifert, \href{https://doi.org/10.1007/BF00758075}{Gen. Relativ. Gravit. 5 (1974) 61}.

\bibitem{uni3}B. Carter \href{ttps://doi.org/10.1103/PhysRevLett.26.331}{Phys. Rev. Lett. 26 (1971) 331}.

\bibitem{uni4}D. C. Robinson, \href{https://doi.org/10.1103/PhysRevLett.34.905}{Phys. Rev. Lett. 34 (1975) 905 .}

\bibitem{uni5}P. T. Chru\'{s}ciel and J. L. Costa, \href{http://www.numdam.org/item/AST_2008__321__195_0/}{Ast\'{e}risque 321 (2008) 195}. 
\bibitem{Bekenstein}
J.D. Bekenstein, \href{https://doi.org/10.1007/BF02757029}{Lett. Nuovo Cimento 4 (1972) 737}.
\bibitem{Bekenstein2}
J. D. Bekenstein, \href{https://link.aps.org/doi/10.1103/PhysRevD.7.2333}{Phys. Rev. D 7 (1973) 2333}.
\bibitem{Hawking}
S.W. Hawking, \href{https://doi.org/10.1038/248030a0}{Nature 248 (1974) 30}.
\bibitem{Hawking2}
S.W. Hawking, \href{https://doi.org/10.1007/BF02345020}{Commun. Math. Phys 43 (1975) 199}.
\bibitem{Hawking3}
S.W. Hawking, \href{https://link.aps.org/doi/10.1103/PhysRevD.13.191}{Phys. Rev. D 13 (1976) 191}.
\bibitem{kastor} D. Kastor, S. Ray and J. Traschen, \href{https://doi.org/10.1088/0264-9381/26/19/195011}{Class. Quantum Grav. 26 (2009) 195011}.
\bibitem{MM}
M. M. Caldarelli, G. Cognola and D. Klemm, \href{http://dx.doi.org/10.1088/0264-9381/17/2/310}{Class. Quantum Grav. 17 (2000) 399}. 
\bibitem{Dolan}
B. P. Dolan, \href{http://dx.doi.org/10.1088/0264-9381/28/12/125020}{Class. Quantum Grav. 28 (2011) 125020}.
\bibitem{Dolan2}
B. P. Dolan, \href{http://dx.doi.org/10.1088/0264-9381/28/23/235017}{Class. Quantum Grav. 28 (2011) 235017}.
\bibitem{Dolan3}
B. P. Dolan, \href{https://link.aps.org/doi/10.1103/PhysRevD.84.127503}{Phys. Rev. D 84 (2011) 127503}.
\bibitem{Cvetic2}
M. Cveti\v{c}, G. Gibbons, D. Kubiz\v{n}\'{a}k and C. Pope, \href{https://link.aps.org/doi/10.1103/PhysRevD.84.024037}{Phys. Rev. D 84 (2011) 024037}.
\bibitem{LuPang}
H. L$\ddot{\text{u}}$, Yi Pang, C. N. Pope and J. F. V\'{a}zquez-Poritz, \href{https://link.aps.org/doi/10.1103/PhysRevD.86.044011}{Phys. Rev. D 86 (2012) 044011}.
\bibitem{revmann1}D. Kubiz\v{n}\'{a}k, R. B. Mann and M. Teo, \href{https://doi.org/10.1088/1361-6382/aa5c69}{Class. Quantum Grav. 34 (2017) 063001 }.
\bibitem{bhc}D. Kubiz\v{n}\'{a}k and R. B. Mann,  \href{https://doi.org/10.1139/cjp-2014-0465}{Can. J. Phys. 93 (2014) 999}.
\bibitem{ext1}
B. P. Dolan, Open Questions in Cosmology , \href{http://dx.doi.org/10.5772/52455}{10.5772/52455 (IntechOpen, 2012)}.
\bibitem{ext2}
D. Kubiz\v{n}\'{a}k and R. B. Mann, \href{https://doi.org/10.1007/JHEP07(2012)033}{J. High Energy Phys. 2012 (2012) 33}
\bibitem{rongPV}R.G. Cai, L.M. Cao, L. Li, and R.Q. Yang, \href{https://doi.org/10.1007/JHEP09(2013)005}{J. High Energy Phys. 09 (2013) 005}
\bibitem{gunasekaran}
S. Gunasekaran, D. Kubiznak and R. B. Mann, \href{https://doi.org/10.1007/JHEP11(2012)110}{J. High Energy Phys. 11 (2012) 110}.
\bibitem{ext3}
Jie-Xiong Mo, Gu-Qiang Li, Shan-Quan Lan and Xiao-Bao Xu, \href{https://doi.org/10.1103/PhysRevD.98.124032}{Phys. Rev. D 98 (2018) 124032}.
\bibitem{ext4}
A. Anabalón, F. Gray, R. Gregory et al., \href{https://doi.org/10.1007/JHEP04(2019)096}{J. High Energ. Phys. 4  (2019) 96}.
\bibitem{ext5}
Yan-Gang Miao and Zhen-Ming Xu, \href{https://doi.org/10.1103/PhysRevD.98.084051}{Phys. Rev. D 98 (2018) 08405}.
\bibitem{mald}
J. M. Maldacena,  \href{https://doi.org/10.1023/A:3A1026654312961}{Adv. Theor. Math. Phys. 2 (1998) 231}.
\bibitem{gub}S.S. Gubser, I.R. Klebanov, A.M. Polyakov, \href{https://doi.org/10.1016/S0370-2693(98)00377-3}{ 	Phys. Lett. B 428 (1998) 105}.
\bibitem{revads} Jens L. Petersen, \href{https://doi.org/10.1142/S0217751X99001676}{ 	Int. J. Mod. Phys. A 14 (1999) 3597}.
\bibitem{wit}
E. Witten, \href{https://dx.doi.org/10.4310/ATMP.1998.v2.n3.a3}{Adv. Theor. Math. Phys. 2 (1998) 505}.
\bibitem{horo}S. A. Hartnoll, C. P. Herzog, and G. T. Horowitz, \href{https://doi.org/10.1103/PhysRevLett.101.031601}{Phys. Rev. Lett. 101 (2008) 031601}.
\bibitem{cv}
C. V. Johnson, \href{https://doi.org/10.1088/0264-9381/31/20/205002}{Class. Quantum Grav. 31 (2014) 205002}.
\bibitem{karch}
A. Karch and B. Robinson, \href{https://doi.org/10.1007/JHEP12(2015)073}{ J. High Energy Phys. 12 (2015) 073}.
\bibitem{visser}
M. R. Visser, \href{https://doi.org/10.1103/PhysRevD.105.106014}{Phys. Rev. D 105 (2022) 106014}.
\bibitem{nss}N. Kumar, S. Sen, and S. Gangopadhyay,  \href{https://doi.org/10.1103/PhysRevD.106.026005}{Phys. Rev. D 106 (2022) 026005}.
\bibitem{Born}
M. Born and L. Infeld, \href{https://doi.org/10.1098/rspa.1934.0059}{Proc. Roy. Soc. Lond. A 144 (1934) 425}.
\bibitem{mann2}
Cong, W., Kubiz\v{n}\'{a}k, D., Mann, R.B. et al. \href{ https://doi.org/10.1007/JHEP08(2022)174}{ J. High Energ. Phys. 2022, 174 (2022)}
\bibitem{DL}
D.L.Wiltshire, \href{https://doi.org/10.1016/0370-2693(86)90681-7}{Phys. Lett. B 169 (1986) 36}.
\bibitem{Smarr} 
L. Smarr, \href{https://doi.org/10.1103/PhysRevLett.30.71}{Phys. Rev. Lett. 30 (1973) 521}.
\bibitem{Gauss1}R. C. Myers and J. Z. Simon, \href{https://doi.org/10.1103/PhysRevD.38.2434}{Phys. Rev. D 38 (1988) 2434}.
\bibitem{Gauss11}
Rong-Gen Cai, \href{https://doi.org/10.1103/PhysRevD.65.084014}{Phys.Rev.D 65 (2002) 084014}.
\bibitem{Gauss2}
Kumar, N., Gangopadhyay, S, \href{https://doi.org/10.1007/s10714-021-02808-0}{ Gen. Relativ. Gravit. 53 (2021) 35}.
\bibitem{Gauss3}
T. Clunan, S. F. Ross and D. J. Smith, \href{https://doi.org/10.1088/0264-9381/21/14/009}{Class. Quantum Grav. 21 (2004) 3447}
\bibitem{Gauss5}
Xu, W., Xu, H. and Zhao, L., \href{https://doi.org/10.1140/epjc/s10052-014-2970-8}{Eur. Phys. J. C 74 (2014) 2970.}
\bibitem{DolanBP}
B. P. Dolan, \href{https://doi.org/10.1007/JHEP10(2014)179}{J. High Energy Phys. 10 (2014) 179}.
\bibitem{Kastor2}
D. Kastor, S. Ray, and J. Traschen, \href{https://doi.org/10.1007/JHEP11(2014)120}{J. High Energy Phys. 11 (2014) 120}.
\bibitem{Zhang1}
J.-L. Zhang., R.-G. Cai, and H. Yu., \href{https://doi.org/10.1007/JHEP02(2015)143}{J. High Energy Phys. 02 (2015) 143}.
\bibitem{Zhang2}
J.-L. Zhang., R.-G. Cai, and H. Yu., \href{https://link.aps.org/doi/10.1103/PhysRevD.91.044028}{Phys. Rev. D 91 (2015) 044028}.
\bibitem{DolanBP2}
B. P. Dolan, \href{ https://doi.org/10.3390/e18050169}{Entropy 18 (2016) 169}.
\bibitem{McCarthy}
F. McCarthy, D. Kubiz\v{n}\'{a}k, and R. B. Mann, \href{https://doi.org/10.1007/JHEP11(2017)165}{J. High Energy Phys. 11 (2017) 165}.
\bibitem{Jhonson2}
C. V. Johnson, V. L. Martin, and A. Svesko, \href{https://link.aps.org/doi/10.1103/PhysRevD.101.086006}{Phys. Rev. D 101 (2020) 086006}.
\end{thebibliography}
\end{document}